\def\year{2018}\relax
\documentclass[letterpaper]{article} 
\usepackage{aaai18}  
\usepackage{times}  
\usepackage{helvet}  
\usepackage{courier}  
\usepackage{url}  
\usepackage{graphicx}  
\expandafter\def\expandafter\UrlBreaks\expandafter{\UrlBreaks
  \do\a\do\b\do\c\do\d\do\e\do\f\do\g\do\h\do\i\do\j%
  \do\k\do\l\do\m\do\n\do\o\do\p\do\q\do\r\do\s\do\t%
  \do\u\do\v\do\w\do\x\do\y\do\z\do\A\do\B\do\C\do\D%
  \do\E\do\F\do\G\do\H\do\I\do\J\do\K\do\L\do\M\do\N%
  \do\O\do\P\do\Q\do\R\do\S\do\T\do\U\do\V\do\W\do\X%
  \do\Y\do\Z}
\usepackage{floatrow}
\usepackage{multirow}
\usepackage[inline]{enumitem}
\DeclareFloatFont{small}{\small}
\begin{document}
%
 \title{Sampling the News Producers: A Large News and Feature Data Set for 
 the Study of the Complex Media Landscape}

\author{ 
Benjamin D. Horne, Sara Khedr, and Sibel Adal{\i} \\
Rensselaer Polytechnic Institute, Troy, New York, USA\\
\{horneb, khedrs, adalis\}@rpi.edu\\
}

\maketitle
 \begin{abstract}
The complexity and diversity of today's media landscape 
provides many challenges for researchers studying news producers.
These producers use many different strategies to get their
message believed by readers through the
writing styles they employ, by repetition across different media 
sources with or without attribution, as well as other mechanisms
that are yet to be studied deeply. To better facilitate 
systematic studies in this area, we present a large political news data set, containing over 136K news articles, from 92 news sources, collected over 7 months of 2017. These news sources are carefully chosen to include well-established and mainstream sources, maliciously fake sources, satire sources, and hyper-partisan political blogs. In addition to each article we compute 130 content-based and social media engagement features drawn from a wide range of literature on political bias, persuasion, and misinformation. With the release of the data set, we also provide the source code for feature computation. In this paper, we discuss the first release of the data set and demonstrate 4 use cases of the data and features: news characterization, engagement characterization, news attribution and content copying, and discovering news narratives. 
 \end{abstract}

\section{Introduction}
The complexity and diversity of today's media landscape provides many challenges for researchers studying news. In this paper, we introduce 
a broad news benchmark data set, called the NEws LAndscape (NELA2017) data set, to facilitate 
the study of many problems in this domain. The data set 
includes articles on U.S. politics 
from a wide range of news sources that includes
well-established news sources, satire news sources, hyper-partisan
sources (from both ends of the political spectrum), as well as, sources that have been known to distribute maliciously
fake information.
At the time of writing, this data set contains 136K news articles from 92 sources between April 2017 and October 2017.

As news producers and distributors can be established quickly with 
relatively little effort, there is limited prior data on the reliability 
of some of many sources, even though the information they provide can end up 
being widely disseminated due to algorithmic and social 
filtering in social media.
It has been argued that the traditionally slow 
fact-checking process and journalistically 
trained ``gatekeepers''are insufficient to counteract the potentially damaging effect
these sources have on the public~\cite{mele2017combating}~\cite{buntain2017automatically}. As a result, there is a great deal of early
research in automatically identifying different writing
styles and persuasion techniques employed by news sources~\cite{popat2016credibility}~\cite{potthast2017stylometric}~\cite{horne2017just}~\cite{chakraborty2016stop} ~\cite{singhania20173han}. Hence, a broad data set including many different types of
sources is especially useful in further refining these methods. To this end, we include 130 
content-based features for each article, in
addition to the article meta-data and full text. The 
feature set contains almost all the features used in the related literature,
such as identifying misinformation, political bias, clickbait, and satire. Furthermore, we include Facebook engagement statistics for each article (the number of shares, comments, and reactions).

While much of recent research has focused on automatic news characterization methods, there are many other news publishing behaviors that are not well-studied. For instance, there are many  sources that have been in existence for a long time.
These sources enjoy a certain level of trust by their audience, sometimes despite 
their biased and misleading reporting, or potentially because of it. 
Hence, trust for sources
and content cannot be studied independently. 
While misinformation in news has attracted 
a lot of interest lately, it is important to note that many sources mix true and false 
information in strategic ways to not only to distribute false information, but also 
to create mistrust for other sources. 
This mistrust and uncertainty may be accomplished by writing specific narratives and having other similar
sources copy that information verbatim~\cite{lytvynenko}. In some cases, sources may copy information 
with the intention to misrepresent it and undermine its reliability. In other cases, a source may copy information to gain credibility itself. Similarly, the coverage of topics in sources
can be highly selective or may 
include well-known conspiracy theories. Hence, it may be important to study a source's output over time and compare it to other sources publishing news in the same time frame. This can sometimes be challenging 
as sources are known to remove articles that attract unwanted attention. We have observed this behavior with 
many highly shared false articles during the 2016 U.S. election.   

These open research problems are the primary reasons we have created the NELA2017 data set.
Instead of concentrating on specific events or specific
types of news, this data set
incorporates all political news production from a diverse group of 
of sources over time.
While many news data sets have been published, none of them have the broad
range of sources and time frame that our data set offers. Our hope is that
our data set can help serve as a starting point for many exploratory 
news studies, and provide a better, shared insight into misinformation tactics. Our aim is to continuously update this data set, expand it with new sources and features, as well as maintain completeness over time.

In the rest of the paper, we describe the data set in detail and provide 
a number of motivating use cases.
The first describe how we can characterize the news sources using the
features we have provided. In the second, we show
how social media engagement differs across groups
sources. We then illustrate content copying behavior among the sources and 
how the sources covered different narratives around two events.

\section{Related Work}
There are several recent news data sets, specifically focused on fake news. 
These data sets include the following.

{\bf Buzzfeed 2016} contains a sample of 1.6K fact-checked news articles from mainstream, fake, and political blogs  shared on Facebook during the 2016 U.S. Presidential Election~\footnote{\url{github.com/BuzzFeedNews/2017-12-fake-news-top-50}}. It was later enhanced with meta data by Potthast et al.~\cite{potthast2017stylometric}. This data set is useful for understanding the false news spread during the 2016 U.S. Presidential Election, but it is unknown how generalizable results will be over different events. 
{\bf LIAR} is a fake news benchmark data set of 12.8K hand-labeled, fact-checked short statements from \url{politifact.com}~\cite{wang2017liar}. This data set is much larger than many previous fake news data sets, but focuses on short statements rather than complete news articles or sources.
{\bf NECO 2017} contains a random sample of three types of news during 2016: fake, real, and satire. Each source was hand-labeled using two online lists. It contains a total of 225 articles~\cite{horne2017just}. While the ground truth is reasonably based, the data set is very small and time-specific.
{\bf BS Detector} contains approximately 12K ``fake news" articles collected using the browser extension BS Detector which labels news based on a manually compiled source dictionary (\url{http://bsdetector.tech/}) and is publicly available on \url{kaggle.com}. The reliability of these lists are unknown.

Additionally, there are much larger, general news data sets that are are focused on events, topics, and location. These include the following.
{\bf GDELT}  contains a wide range of online publications, including news and blogs, in over 100 languages. The collection is based on world events, focusing on location, temporal, and network features. GDELT provides a useful visual knowledge graph that indexes images and visuals used in news. While this data set provides news data over an extended period of time, it is focused on news surrounding external events, and may not capture many ``fake" news sources. In addition, Kwak and An~\cite{kwak2016revealing} point out that there is concern as to how biased the GDELT data set is as it does not always align with other event based data sets. 
{\bf Unfiltered News}  (\url{unfiltered.news}) is a service built by Google Ideas and Jigsaw to address filter bubbles in online social networks. Unfiltered News indexes news data for each country based on mentioned topics. This data set does not focus on raw news articles or necessarily false news, but on location-based topics in news, making it extremely useful for analyzing media attention across time and location. Data from Unfiltered News is analyzed in ~\cite{an2017convergence}.

There are many more data sets that focus on news or claims in social networks.
{\bf CREDBANK} is a crowd sourced data set of 60 million tweets between October 2015 and February 2016. Each tweet is associated to a news event and is labeled with credibility by Amazon Mechanical Turkers~\cite{mitra2015credbank}. This data set does not contain raw news articles, only news article related tweets.
{\bf PHEME} is a data set similar to CREDBANK, containing tweets surrounding rumors. The tweets are annotated by journalist~\cite{zubiaga2016analysing}. Once again, this data set does not contain raw news articles, but focused on tweets spreading news and rumors. Both PHEME and CREDBANK are analyzed in~\cite{buntain2017automatically}.
{\bf Hoaxy} is an online tool that visualizes the spread of claims and related fact checking~\cite{Shao:2016:HPT:2872518.2890098}. Claim related data can be collected using the Hoaxy API. Once again, data from this tool is focused on the spread of claims (which can be many things: fake news article, hoaxes, rumors, etc.) rather than news articles themselves.

Other works use study-specific data sets collected from a few sources. Some of these data sets are publicly available. Piotrkowicz et al. use 7 months of news data collected from The Guardian and The New York Times to assess headline structure's impact on popularity~\cite{piotrkowicz2017headlines}. Reis et al. analyze sentiment in 69K headlines collected from The New York Times, BBC, Reuters, and Dailymail~\cite{reis2015breaking}. Qian and Zhai collect news from CNN and Fox News to study unsupervised feature selection on text and image data from news~\cite{qian2014unsupervised}. Saez-Trumper at al. explore different types of bias in news articles from the top 80 news websites during a two-week period~\cite{saez2013social}. 

There are 3 core issues with these data sets that we address with the NELA2017 data set:

\begin{enumerate}
    \item Small in size and sources - The current data sets that focused on news producers contain very few sources, typically focused on one type of source (mainstream, fake, etc.), and have a small number of data points.
    \item Event specific - Many of the current data sets are focused on small time frames or specific events (ex. 2016 Presidential Election). To ensure current results can be generalized and to track how the news is changing, researchers need data across time and events.
    \item Engagement specific - The majority of these data sets contain only highly engaged or shared articles. While it can be argued that these are the more important data points, they lack the complete picture of news producer behavior. In order to understand how news producers publish, specifically hyper-partisan and malicious sources, researchers need to explore both the viral and the never seen articles produced.
\end{enumerate}

Hence, our goal for the NELA2017 data set is to create a large, near-complete news article data set, across the various types of sources, in hopes of providing a more complete view of how news producers behave.

\begin{table*}
\begin{center}
\fontsize{7.95}{8}\selectfont
\hspace*{-0.0in}\begin{tabular}{|c|c||c|c||c|c||c|c|}
\hline
\textbf{Source} & \textbf{Complete} & \textbf{Source} & \textbf{Complete} & \textbf{Source} &\textbf{Complete} & \textbf{Source} & \textbf{Complete} \\
\hline
AP & 50\%  & Freedom Daily & 100\%  & Observer & 100\%  & Duran & 71\% \\
Activist Post & 100\%  & Freedom Outpost & 100\%  & Occupy Democrats & 93\%  & Fiscal Times & 71\% \\
Addicting Info & 57\%  & FrontPage Mag & 100\%  & PBS & 100\%  & Gateway Pundit & 100\% \\
Alt Media Syn & 78\%  & Fusion & 86\%  & Palmer Report & 50\%  & The Guardian & 100\% \\
BBC & 100\%  & Glossy News & 100\%  & Politicus USA & 100\%  & The Hill & 100\% \\
Bipartisan Report & 100\%   & Hang the Bankers & 72\%  & Prntly & 71\%  & Huffington Post & 100\% \\
Breitbart & 100\%   & Humor Times & 100\%  & RT & 71\%  & The Inquisitr & 100\% \\
Business Insider & 100\%   & Infowars & 100\%  & The Real Strategy & 100\%  & New York Times & 100\% \\
BuzzFeed & 100\%   & Intellihub & 100\%  & Real News Right Now & 100\%  & The Political Insider & 100\% \\
CBS News & 100\%   & Investors Biz Daily & 100\%  & RedState & 100\%  & Truthfeed & 79\% \\
CNBC & 100\%   & Liberty Writers & 100\%  & Salon & 100\%  & The Right Scoop & 100\% \\
CNN & 100\%   & Media Matters & 100\%  & Shareblue & 50\%  & The Shovel & 100\% \\
CNS News & 100\%   & MotherJones & 36\%  & Slate & 100\%  & The Spoof & 100\% \\
Conservative Trib & 100\%   & NODISINFO & 86\%  & Talking Points Memo & 50\%  & TheBlaze & 100\% \\
Counter Current & 100\%   & NPR & 100\%  & The Atlantic & 100\%  & ThinkProgress & 100\% \\
Daily Buzz Live & 86\%   & National Review & 100\% & The Beaverton & 100\%  & True Pundit & 100\% \\
Daily Kos & 100\%  & Natural News & 100\%  & Borowitz Report & 93\%  & Washington Examiner & 100\% \\
Daily Mail &  100\%  & New York Daily & 100\%  & Burrard Street Journal & 86\%  & USA Politics Now & 36\% \\
Daily Stormer & 72\%  & New York Post & 100\%  & The Chaser & 100\%  & USA Today & 100\% \\
Drudge Report & 79\%  & NewsBiscuit & 100\%  & ConservativeTreeHouse & 100\%  & Veterans Today & 100\% \\
Faking News & 100\%  & NewsBusters & 72\%  & D.C. Clothesline & 93\%  & Vox & 100\% \\
Fox News & 86\%  & Newslo & 93\%  & Daily Beast & 100\%  & Waking Times & 100\% \\
World News Politics & 93\%  & Xinhua & 36\%  & Yahoo News & 100\%  & Young Conservatives & 93\% \\
\hline
\end{tabular}
\caption{Approximate completion percentage of all sources in the data set. Since each news source publishes at different rates, we compute completion as having more than 1 article published in each 2 week period of the data set.}\label{sources}
\end{center}
\end{table*}

\section{Data set creation}
In creating our data set, we target a collection of sources to include
both well-established news companies, political blogs, and satire websites
,as well as many alternative news sources that have published 
misinformation in the past or have relatively unknown veracity.
To select these sources, we used a 3-step process: \begin{enumerate*}\item We select well-known sources using Wikipedia lists to capture many mainstream and well-established sources. \item We randomly select sources from the \url{opensources.co} lexicon. OpenSources is expert-curated news source lexicon containing  12 different types of sources: fake, satire, extreme bias, conspiracy, rumor, state, junk science, hate speech, clickbait, unreliable, political, and reliable. This step captures many niche sources and those who have spread fake news in the past. \item We hand select sources cited by previously selected sources (based on reading random articles). \end{enumerate*} This 3rd step provides even more diversity across intentions and political leanings. To ensure that we have a balance of left and right leaning sources, we review selected sources using the crowd-sourced bias-checking service \url{mediabiasfactcheck.com}.

Once we have the set of news sources, we create article scrapers for each source. Each scraper is collects news articles at 12:00pm EST and 9:00pm EST each day. This near real-time collection allows us to maintain news articles that are later deleted, a common practice among maliciously fake new sources. 
Some sources can be collected using standard RSS feed scrapers, while others, especially the less credible sources, need custom web scrapers to collect articles. For news sources with available RSS feeds, we use the Python library feedparser~\footnote{pythonhosted.org/feedparser/}, for news sources with standard HTML structure we use python-goose~\footnote{github.com/grangier/python-goose}, and for news sources with difficult to parse HTML structures, we use a mix of BeautifulSoup~\footnote{www.crummy.com/software/BeautifulSoup/bs4/doc/}, and feedparser to create site specific scrapers. Of the 100 sources selected, there were 8 that our scrapers could not consistently collect, leaving us with 92 sources.

To control for topic, we only collect political news from each source. For the majority of sources, controlling for topic is very easy, as their websites are divided into topic-based feeds. It is important to note that some topic-based feeds are less strict than others, specifically on fake news sites. Thus, in the political news feed, some pseudo-science and odd topic conspiracy articles are mixed in. We choose to collect these occasional off-topic articles as well, as they may provide insight to these fake news sources. 

\noindent Each scraper collects the following information: \\
\indent \textbf{content} - the text from the body of the article \\
\indent \textbf{title} - the text from the title of the article \\
\indent \textbf{source} - the source of the article \\
\indent \textbf{author} - the journalist who wrote the article, if the information is available in the web page meta data \\
\indent \textbf{published} - the UTC time stamp of publication according to the web page \\
\indent \textbf{link} - the url used to scrape the article (RSS feed or web page) \\
\indent \textbf{html} - the full HTML of the article page stored as unicode \\

This information is stored for each article in a JSON dictionary, with keys of the same name as above. 

Using this process, we obtain almost 100\% of the articles produced during the 7 month time period. The approximate completion percentage for each source over the 7 months of collection can be found in Table~\ref{sources}.

\begin{figure*}[h]
\begin{center}
  \begin{tabular}{cc} \\
  \small{(a) Top 10 Most Subjective Writing Style (on average)} & \small{(b) Top 10 Hardest to Read (on average)}\\
    \includegraphics[width=200pt,keepaspectratio=true]{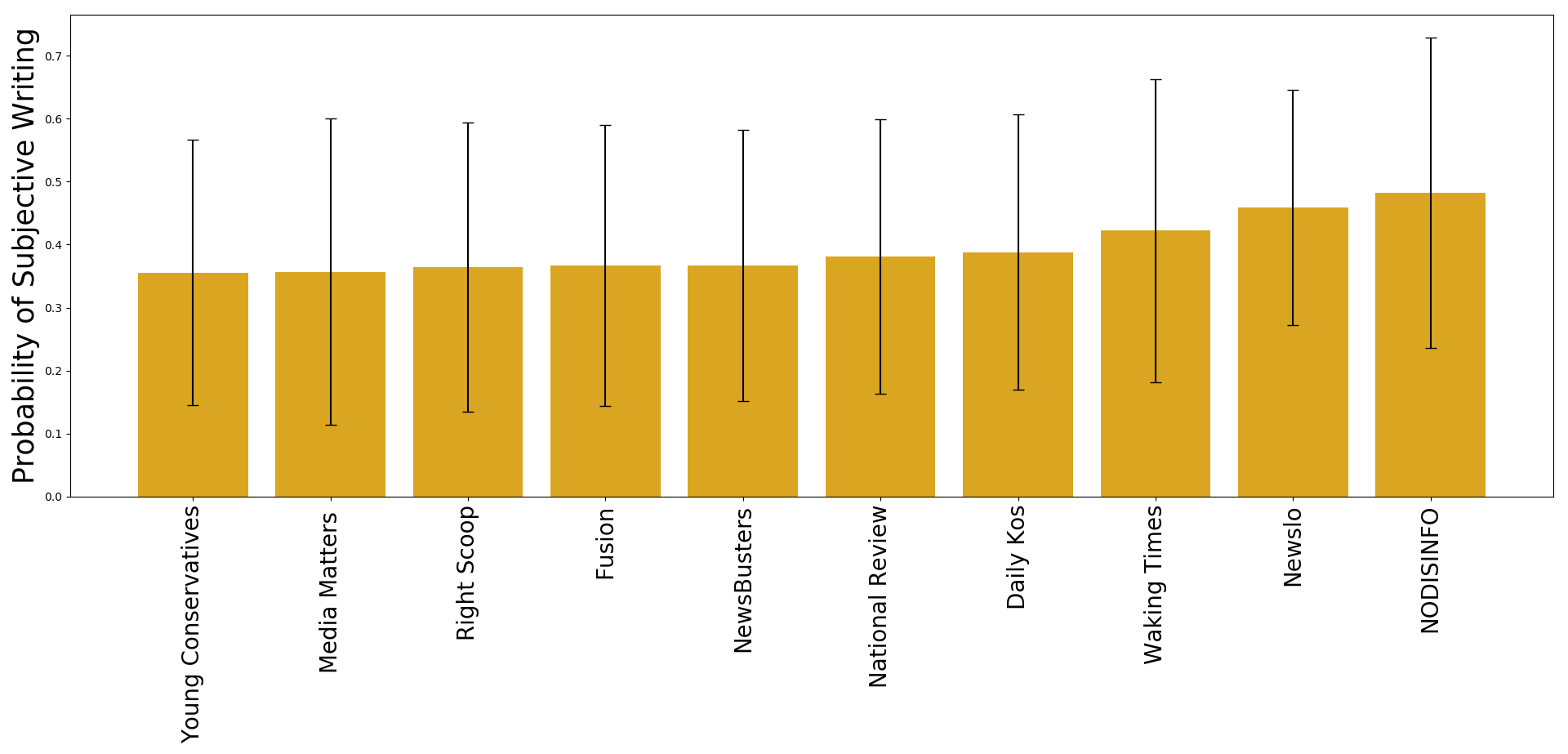}&
    \includegraphics[width=200pt,keepaspectratio=true]{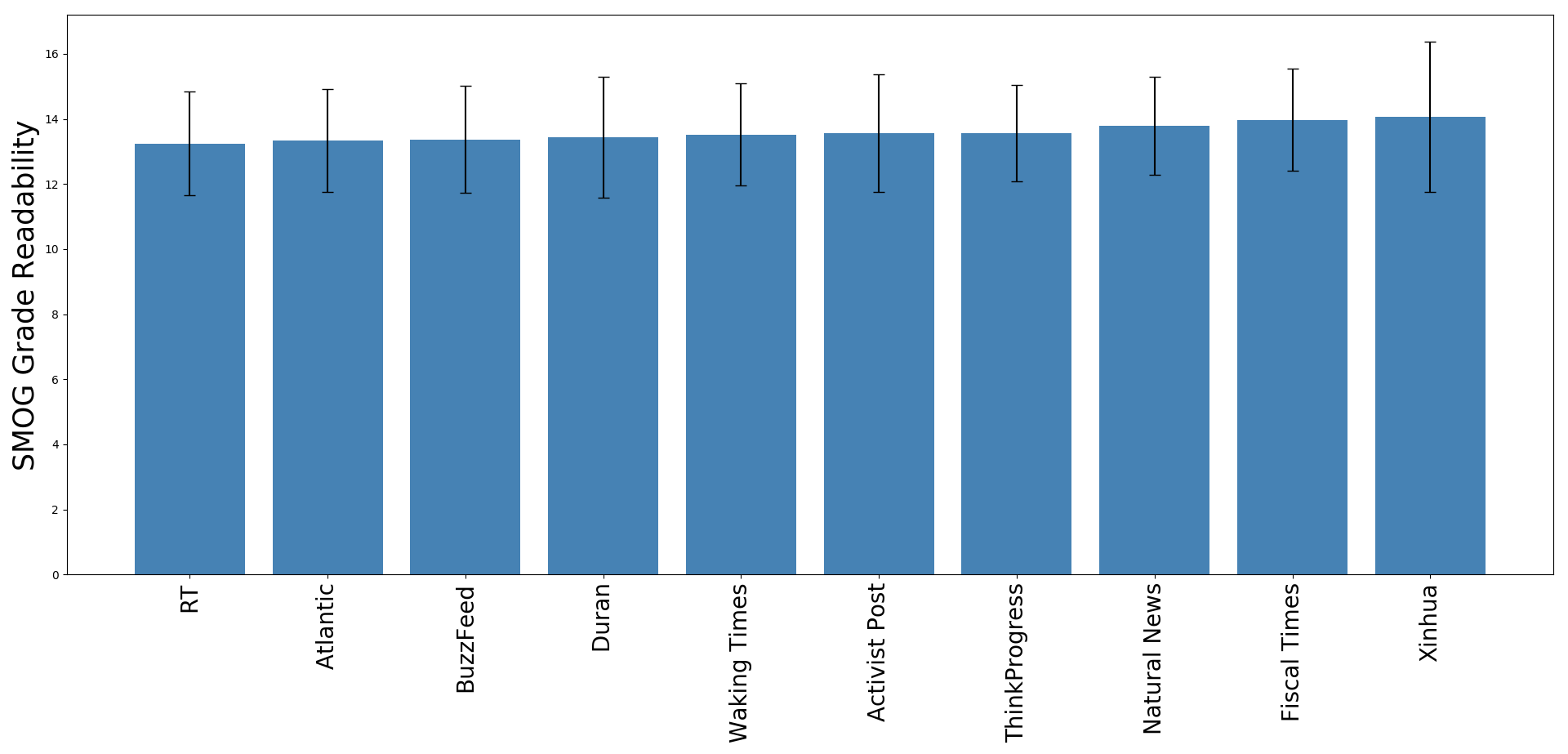}\\
    \small{(c) Top 10 Most Clickbait Titles (\% of articles)} & \small{(d)Top 10 Longest Title (on average)}\\
    \includegraphics[width=200pt,keepaspectratio=true]{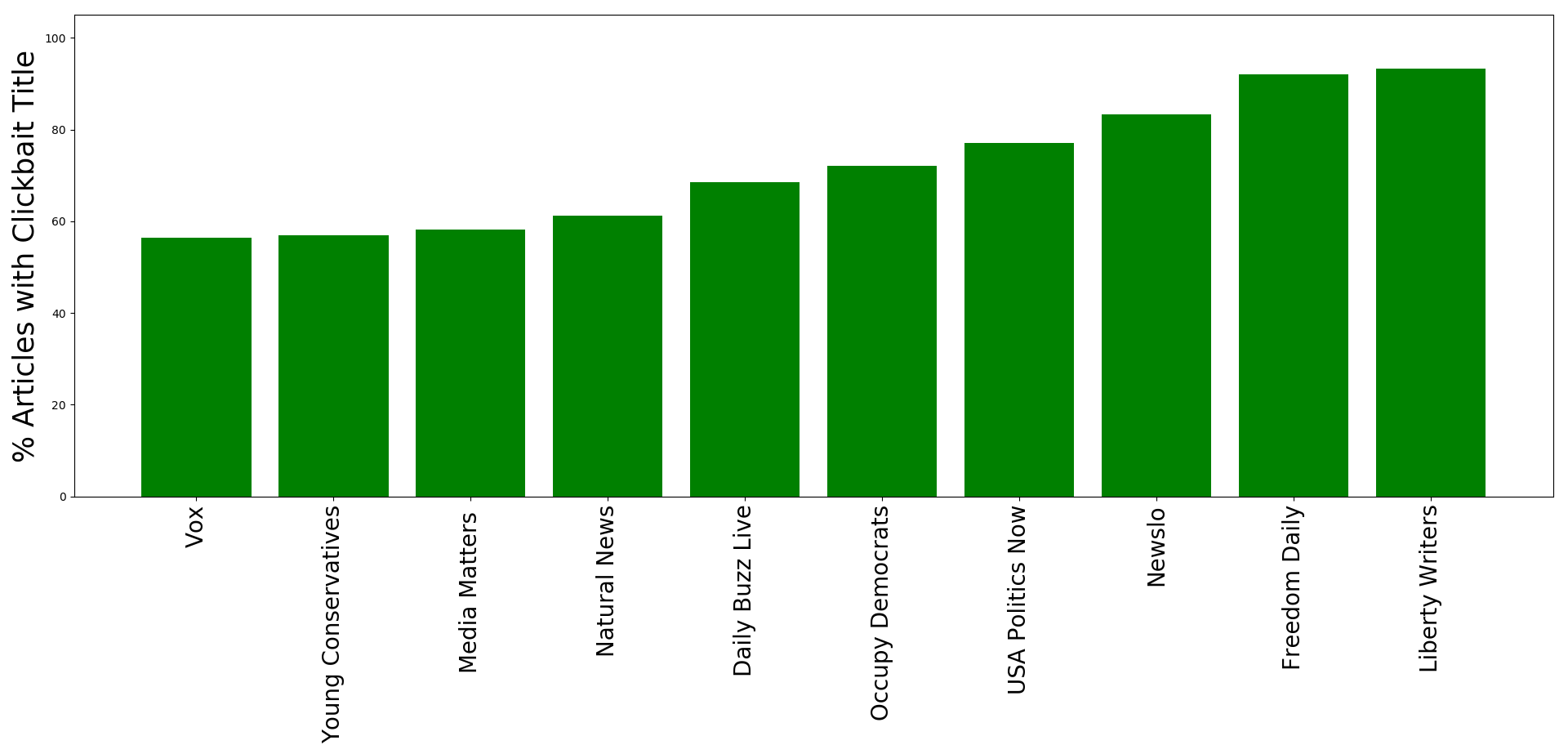}&
    \includegraphics[width=200pt,keepaspectratio=true]{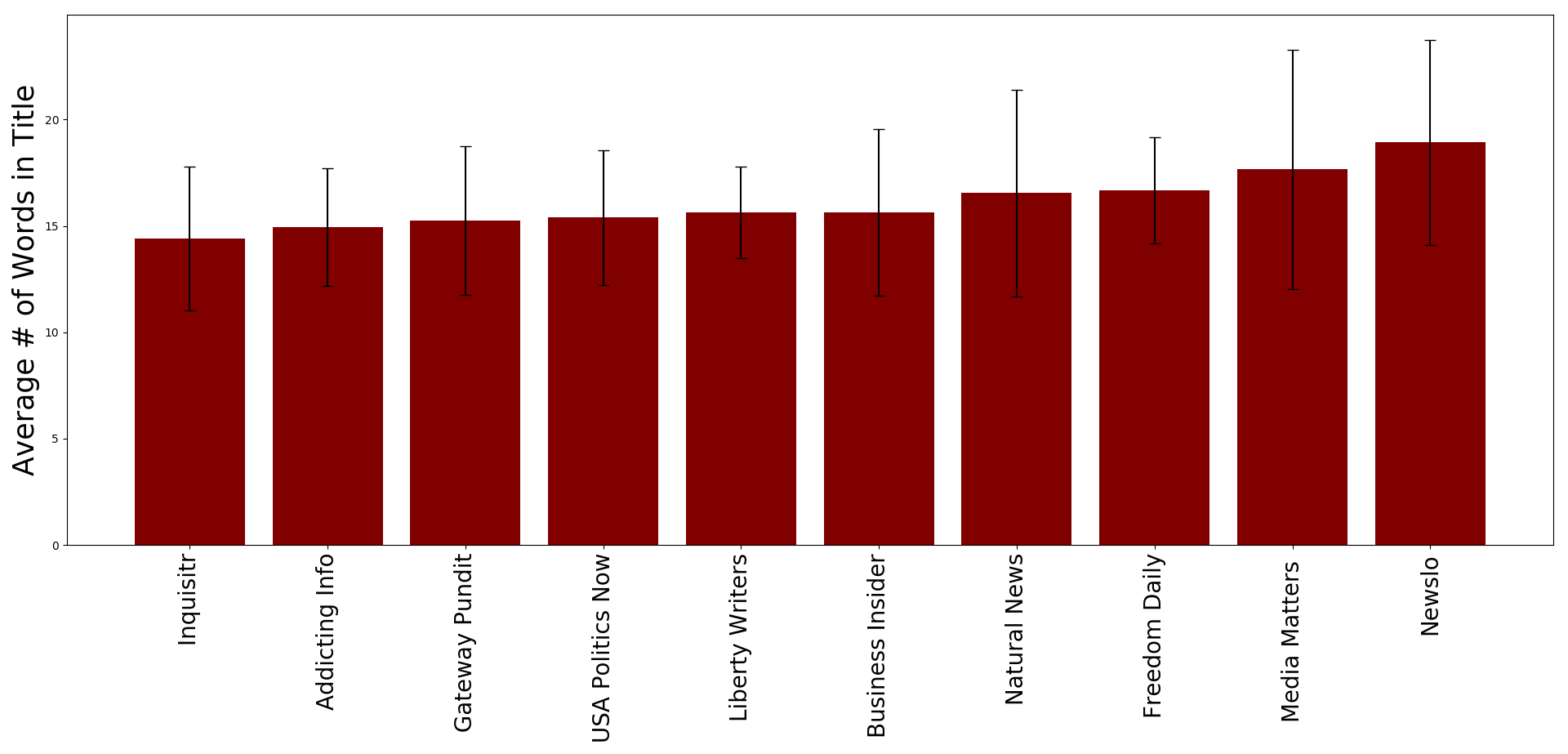}\\
    \small{(e) Top 10 Most Negative Sources (on average)} & \small{(f) Top 10 Most Lexically Redundant Sources (on average)}\\
    \includegraphics[width=200pt,keepaspectratio=true]{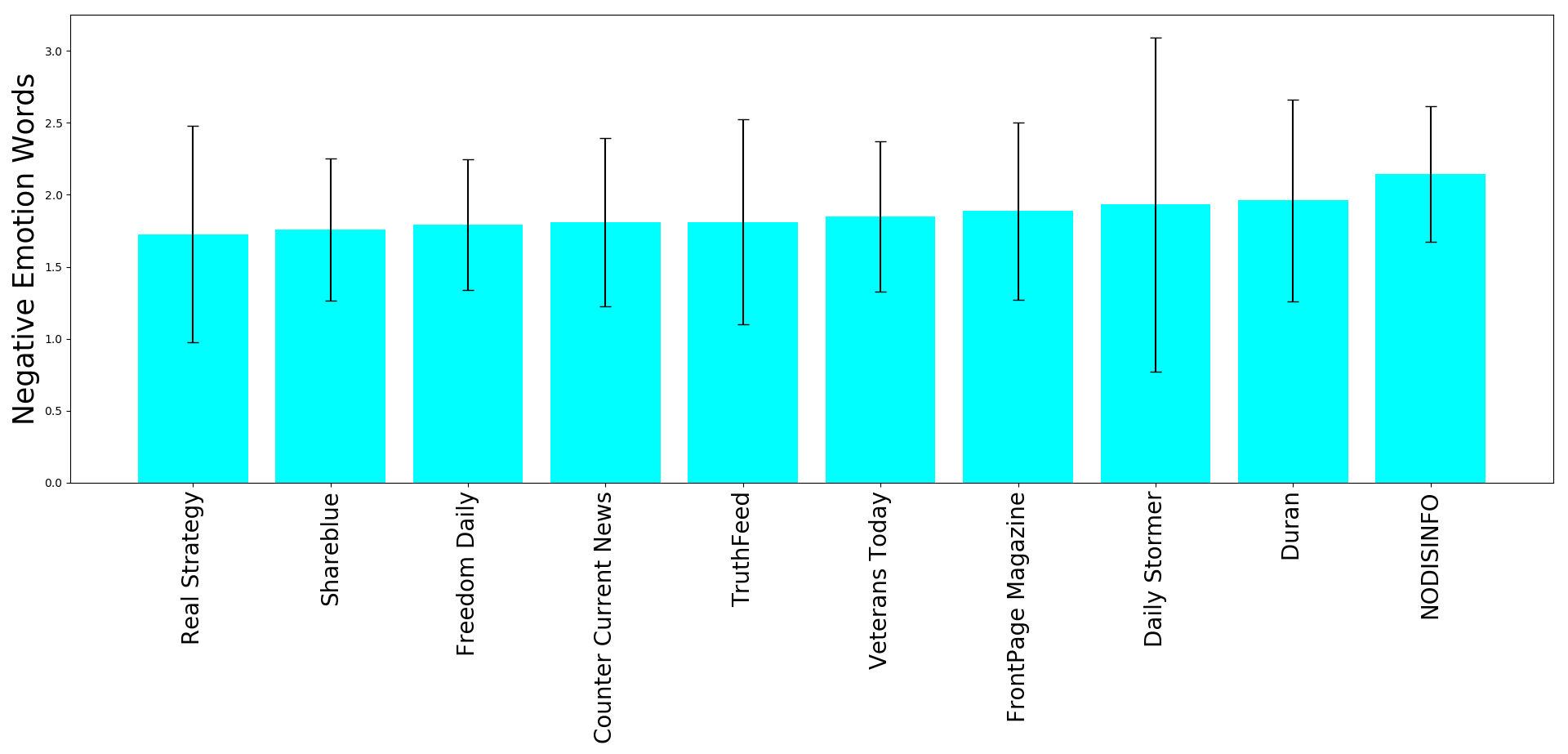}&
    \includegraphics[width=200pt,keepaspectratio=true]{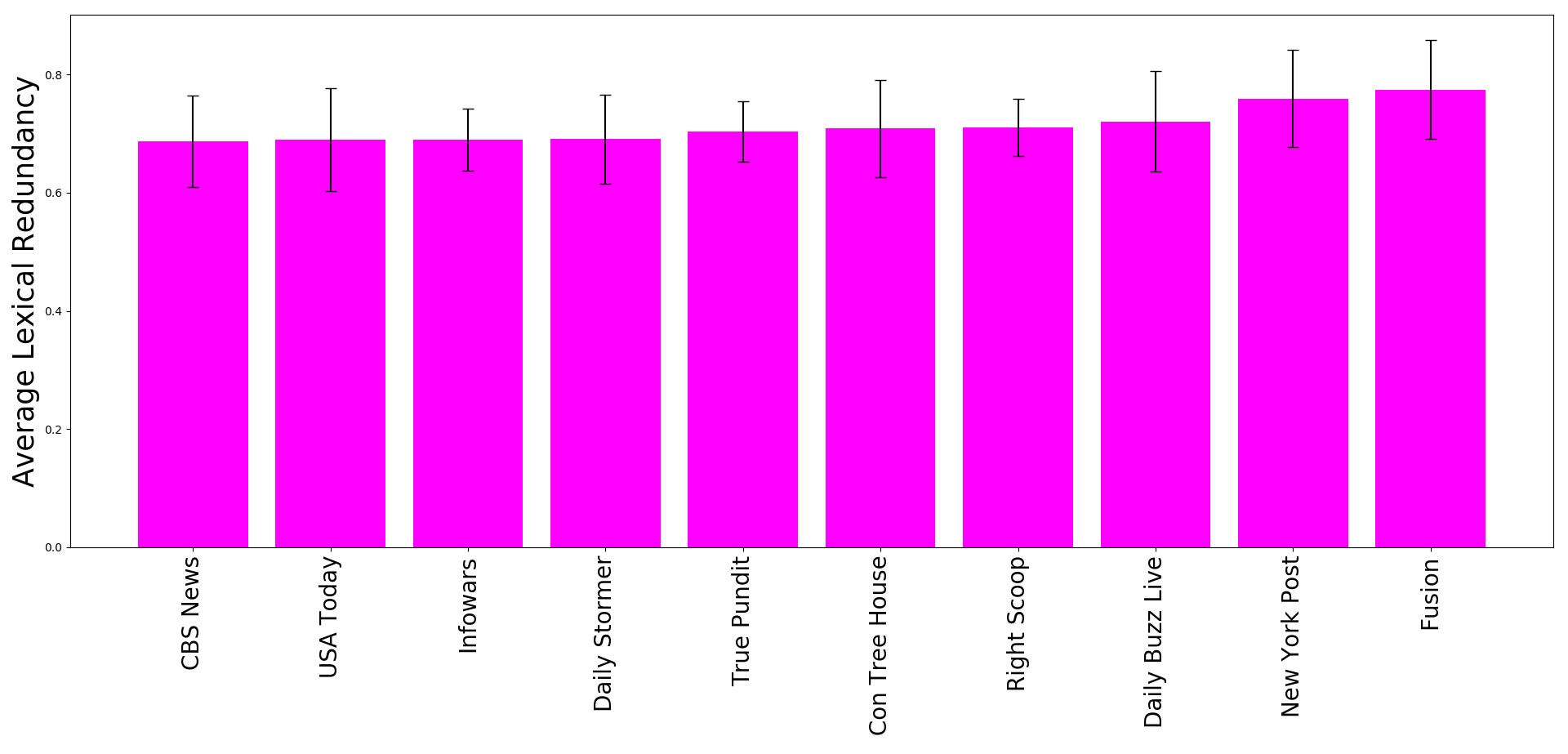}\\
\end{tabular}
\vspace*{-0.2in}  \caption{Top 10 sources for a selection of features.\label{top10}}
\end{center}
\end{figure*}

\begin{figure*}[ht]
\begin{center}
\hspace*{-0.1in}\begin{tabular}{cccc} \\
    \includegraphics[width=115pt,keepaspectratio=true]{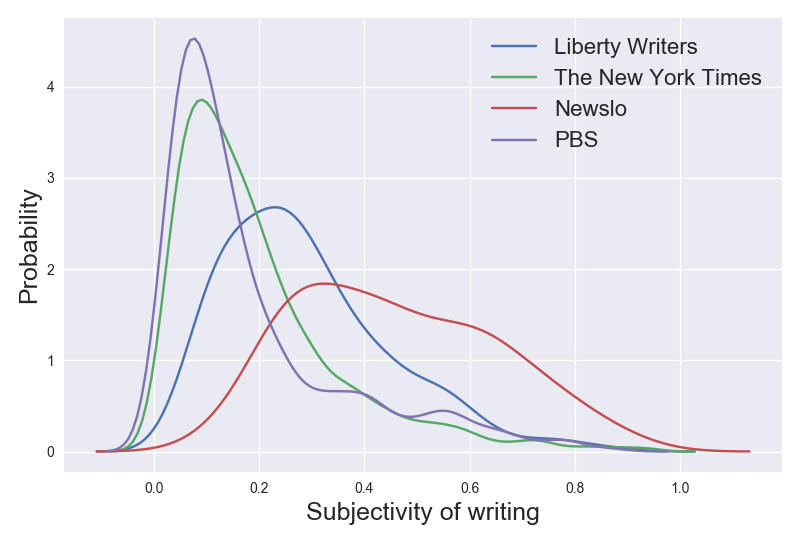}
    & \includegraphics[width=115pt,keepaspectratio=true]{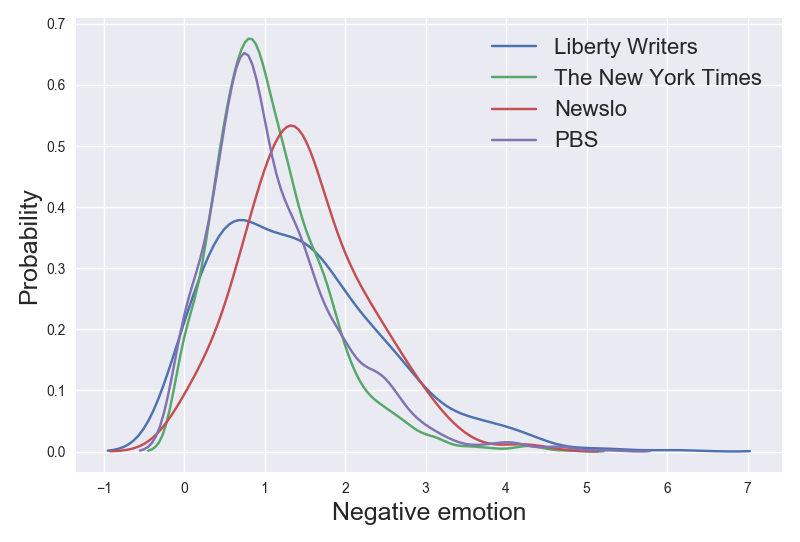}
    &\includegraphics[width=115pt,keepaspectratio=true]{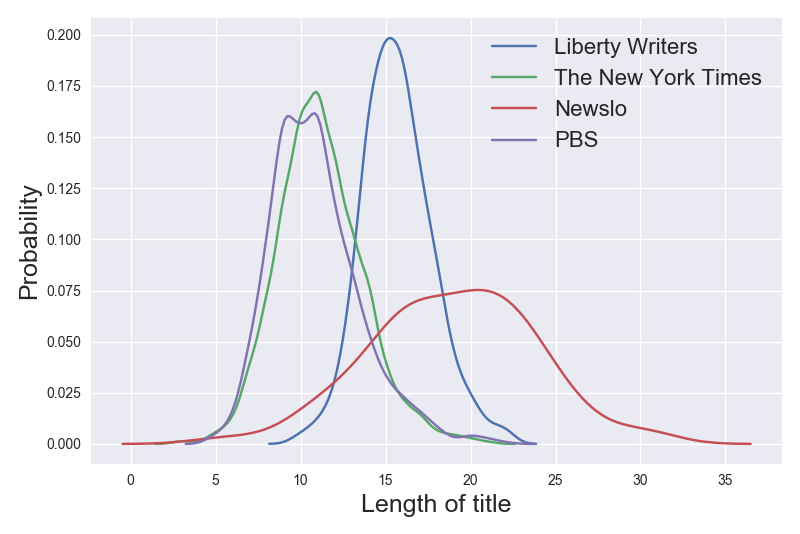}
    &\includegraphics[width=115pt,keepaspectratio=true]{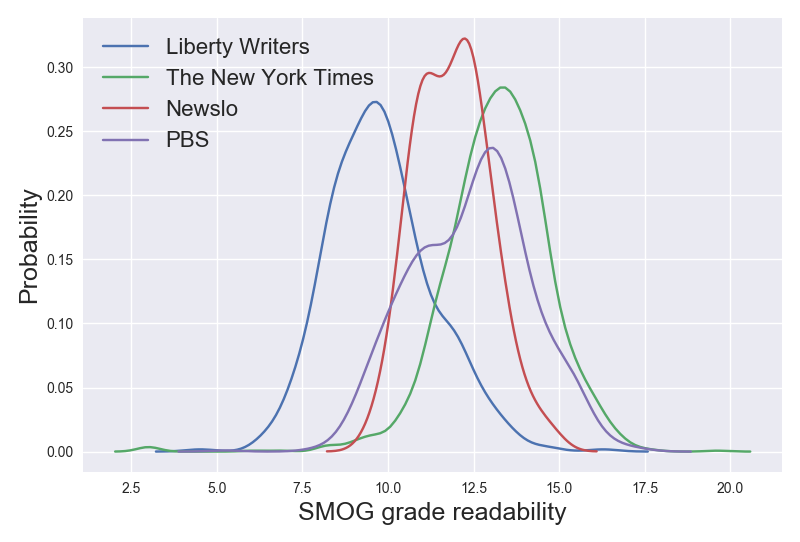} 
\end{tabular}
 \vspace*{-0.2in}  \caption{Feature distributions across different articles from specific
  sources\label{dists}}
\end{center}
\end{figure*}


 \begin{figure*}[th]
\begin{center}
  \hspace*{-0.1in}\begin{tabular}{cccc} \\
    {\small Median} & {\small Max} & {\small Median} & {\small Max} \\
    \includegraphics[width=115pt,keepaspectratio=true]{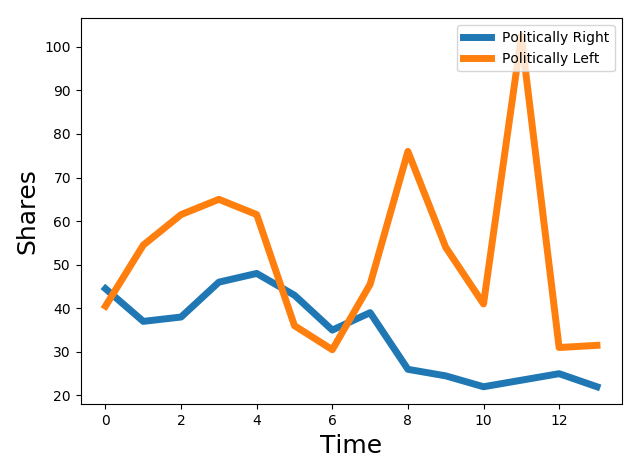}
    & \includegraphics[width=115pt,keepaspectratio=true]{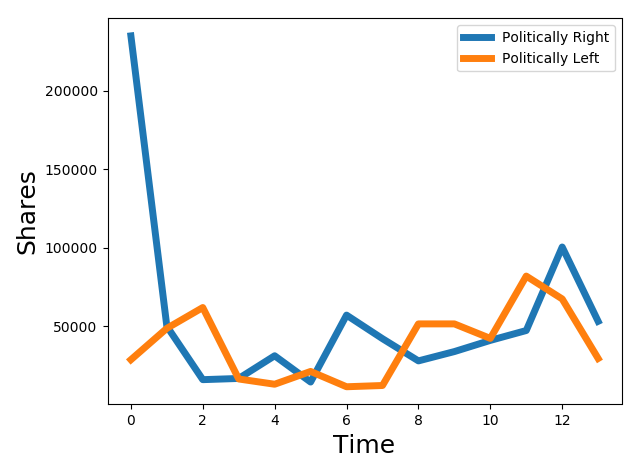} &
        \includegraphics[width=115pt,keepaspectratio=true]{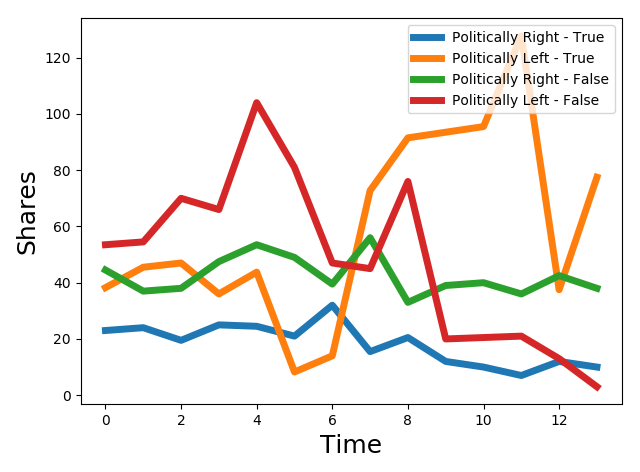}
    & \includegraphics[width=115pt,keepaspectratio=true]{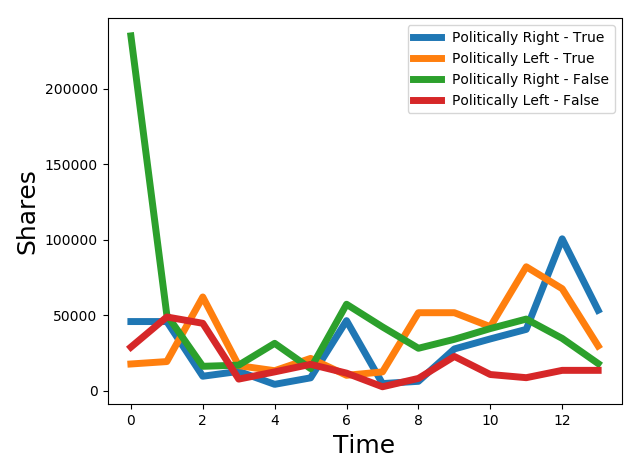} \\
    \multicolumn{2}{c}{(a) Self-Proclaimed Political biased groups}  &
      \multicolumn{2}{c}{(b) Self-Proclaimed Political + previous behavior groups} \\
\end{tabular}
 \vspace*{-0.1in}  \caption{Facebook shares for source groups over time. The median or the max of the shares is measure every two weeks. Hence, 0 is the beginning of April, and 14 is the end of October. \label{fb}}
\end{center}
\end{figure*}

\begin{table*}[thb!]
\centering
\begin{minipage}[t]{3.2in}
\hspace*{-0.15in}\begin{tabular}{p{0.6in}p{2.3in}} 
\bf{Abbr.} &\bf{Description} \\ \hline
    &\\
	POS & normalized count of each part of speech (36 feats)\\ 
	linguistic & \# function words, pronouns, articles, prepositions, verbs, etc. using LIWC lexicons (24 features)\\
	clickbait & clickbait title classification using models built in~\cite{chakraborty2016stop}\\
	\hline
     \multicolumn{2}{c}{\bf(a) Structure Features}  \\\\
     \hline
     sentiment & negative, positive, and neutral sentiment scores from VADER~\cite{hutto2014vader} (3 features)\\
  	emotion & positive, negative, affect, etc. words using LIWC and strong/weak emotion words from lexicons in~\cite{recasens2013linguistic} (13 features) \\
  	Happiness & happiness score using~\cite{mitchell2013geography} Happiness lexicon\\ 
  	&\\
  	\hline 
\multicolumn{2}{c}{\bf(b) Sentiment Features} \\ \\ \hline
Facebook engagement & \# of shares, comments, reactions collected using Facebook API\\
  \hline
  \multicolumn{2}{c}{\bf(c) Engagement Features} \\
  bio & biological processes from LIWC lexicon (5 features)\\
  relativity & motion, time, and space words from LIWC lexicon (4 features)\\
  personal concerns & work, home, leisure, etc. from LIWC lexicon (6 features)\\
  &\\
  \hline
  \multicolumn{2}{c}{\bf(d) Topic-dependent Features} \\
\end{tabular}
  \end{minipage} 
\begin{minipage}[t]{3.2in}
\hspace*{-0.1in}\begin{tabular}{p{0.6in}p{2.3in}}  
\bf{Abbr.} &\bf{Description} \\ \hline
TTR & Type-Token Ratio, also known as lexical diversity or redundancy, computed as $\frac{\# unique words}{total words}$\\
&\\
 FKE & Standard readability measure computed by $0.39 * (\frac{total words}{total sentences}) + 11.8 * (\frac{total syllables}{total words}) - 15.59$\\
 
 SMOG & Standard readability measure computed by $1.0430 * \sqrt{\#polysyllables * \frac{30}{\#sentences}} + 3.1291$\\
 &\\
 wordlen & average \# characters in a word\\
 WC & word count\\
 cogmech & \# cognitive process words (includes cause, insight, etc.) from LIWC lexicons (7 features)\\
 \hline 
 \multicolumn{2}{c}{\bf(e) Complexity Features}  \\ \\ \hline
bias & several bias lexicons from ~\cite{recasens2013linguistic} and ~\cite{mukherjee2015leveraging} (14 features)\\
subjectivity & probability of subjective text using a Naive Bayes classifier trained on 10K subjective and objective sentences from~\cite{pang2004sentimental} used in~\cite{horne2017just}\\
	\hline 
\multicolumn{2}{c}{\bf(f) Bias Features} \\ \\ \hline
  Moral & features based on Moral Foundation Theory~\cite{graham2009liberals} and lexicons used in ~\cite{lin2017acquiring} (10 features)\\
  &\\
  \hline
  \multicolumn{2}{c}{\bf(g) Morality Features} \\ \\
\end{tabular}
\end{minipage}
\caption{\label{tbl:features} Different features implemented on data set. Each feature is compute on the title and body text separately}
\end{table*}

{\bf Feature set creation} 
Next, to facilitate content-based analysis and writing style research on these articles, we compute 130 content-based features and collect 3 Facebook engagement statistics on each news article. These features come from a wide range of literature on false news detection~\cite{potthast2017stylometric}~\cite{horne2017just}~\cite{horne2018accessing}, political bias detection~\cite{recasens2013linguistic}, content popularity~\cite{piotrkowicz2017headlines}~\cite{horne2017identifying}, clickbait detection~\cite{chakraborty2016stop}, and general text characterization~\cite{loper2002nltk}. We break these features down into 7 categories: structure, complexity, sentiment, bias, morality, topic, and engagement. All 130 features are computed on the title and the body text separately, giving us 260 content-based features in total. Due to the wide range of literature these features are borrowed from, some are highly correlated, but all are computed differently. 
To allow researchers even more flexibility, we provide all of the feature code in one easy-to-use Python script. 
All feature code and implementation details are available at: \url{https://github.com/BenjaminDHorne/Language-Features-for-News}. Descriptions of these features can be found in Table~\ref{tbl:features}. Due to lack of space, we will leave major implementation details to the data set and code documentation.

\section{Potential use cases of the NELA2017 data set}
There is a variety of news credibility research strands that can benefit from this data set. In particular, we argue that this data set can not only test the generality of previous results in computational journalism, but also spark research in lesser studied areas. In this section, we present 4 use cases with varying levels of granularity, including: general news source characterization, highly engaged article characterization, content attribution and copying, and analyzing specific news narratives.

\subsection{News Source Characterization}
The most obvious and general use of the NELA2017 data set is news source characterization
and comparison. With the increasing public attention on news sources, many maps of the media
landscape have been offered to show how different sources compare to each other.
Often these maps are based on a subjective evaluation of these sources. Our features make it
possible to draw such comparisons based on algorithms with transparent criteria. 

We first show the top 10 sources
in Figure~\ref{top10} according to their average behavior with respect:
(a) subjectivity based on writing style, (b) grade level readability, 
(c) the clickbait nature of titles, 
(d) length of titles, (e)d negative sentiments expressed, and 
(f) the amount lexical redundancy, i.e. repetition in articles.
Past research shows fake news articles are generally easier to read 
and more repetitive, but are not necessarily clickbait~\cite{horne2017just}. It is also well-studied that
many highly engaged fake articles and conspiracy theories express negative
emotions~\cite{Bessi:2015hg}. All of these previous results are accurately supported by the ranking with our features. For example, the subjectivity accurately captures a
number of highly partisan sources in our list and the clickbait predictions point to
well-known clickbait sources. However, these clickbait sources are not necessarily
among the sources with very long titles or repetitive content. The sources with
highest grade reading include some sources that translate languages (Xinhua) and more niche domain sources (The Fiscal Times). 

Additionally, we also
look at the consistency of sources are over time. 
Sources may show higher variation in 
these distributions due to lack of editorial standards, as well as,
different types of content mixing (made up content or content copied from other sources).
In Figure~\ref{dists}, we show select feature distributions over the full 7 months of data for four news sources: Liberty Writers, Newslo, The New York Times, and PBS. We can clearly see both Liberty Writers and Newslo have very wide distributions, where as The New York Times and PBS have much more narrow distributions, illustrating consistency. These features are not only useful for quick source comparison, but have predictive power in news as shown in prior work~\cite{popat2016credibility}~\cite{horne2017just}. Given our feature set is a superset of all the features from the different literature threads, we expect them to have accuracy as well or better
than those reported. Due to lack of space, we do not provide examples of prediction.


\begin{figure*}[ht] 
\centering
\hspace*{-0.5in}\begin{tabular}{cc}
\small{(a) May 1st-14th 2017} & \small{(b) July 1st-14th 2017}\\
\includegraphics[width=9cm]{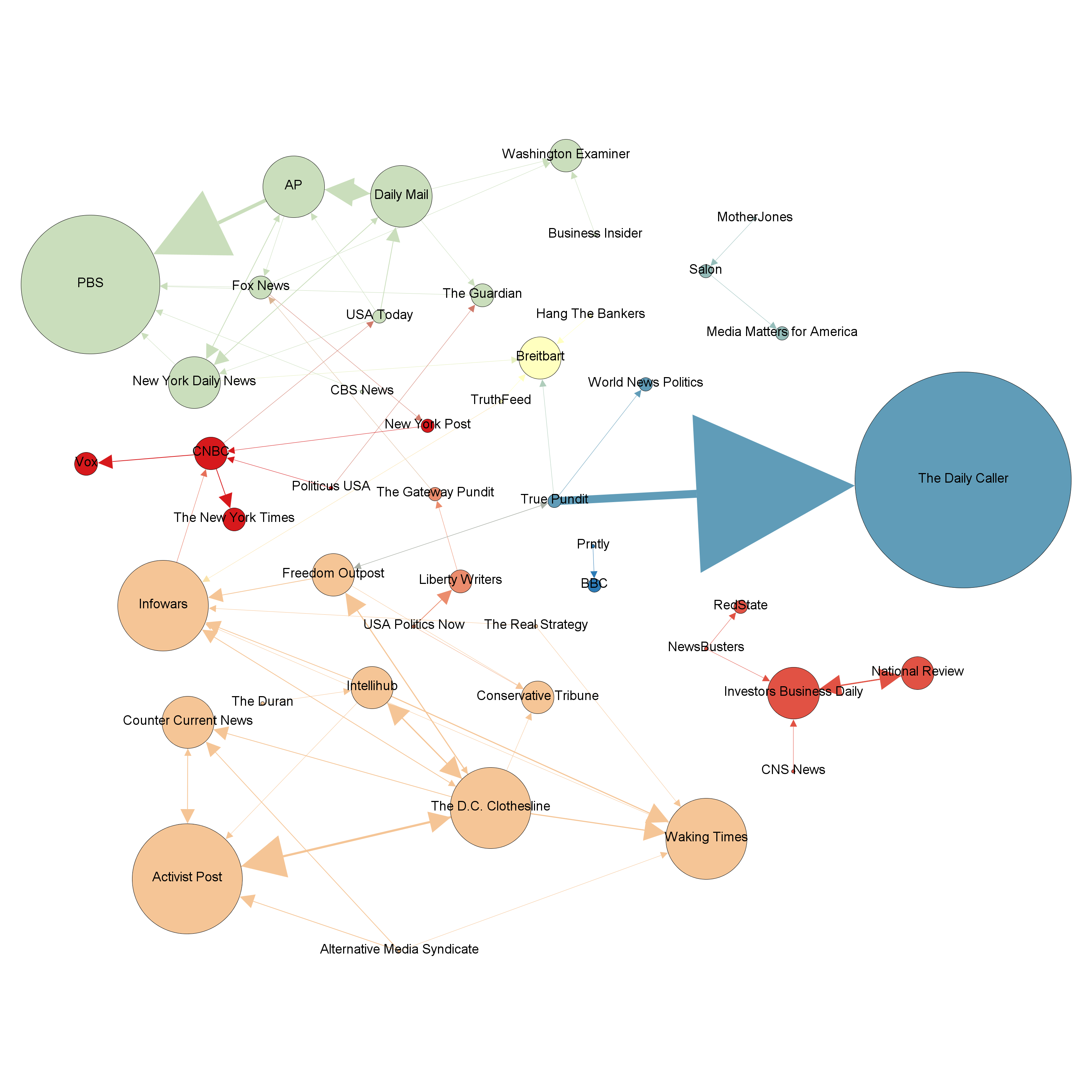}&
\includegraphics[width=9cm]{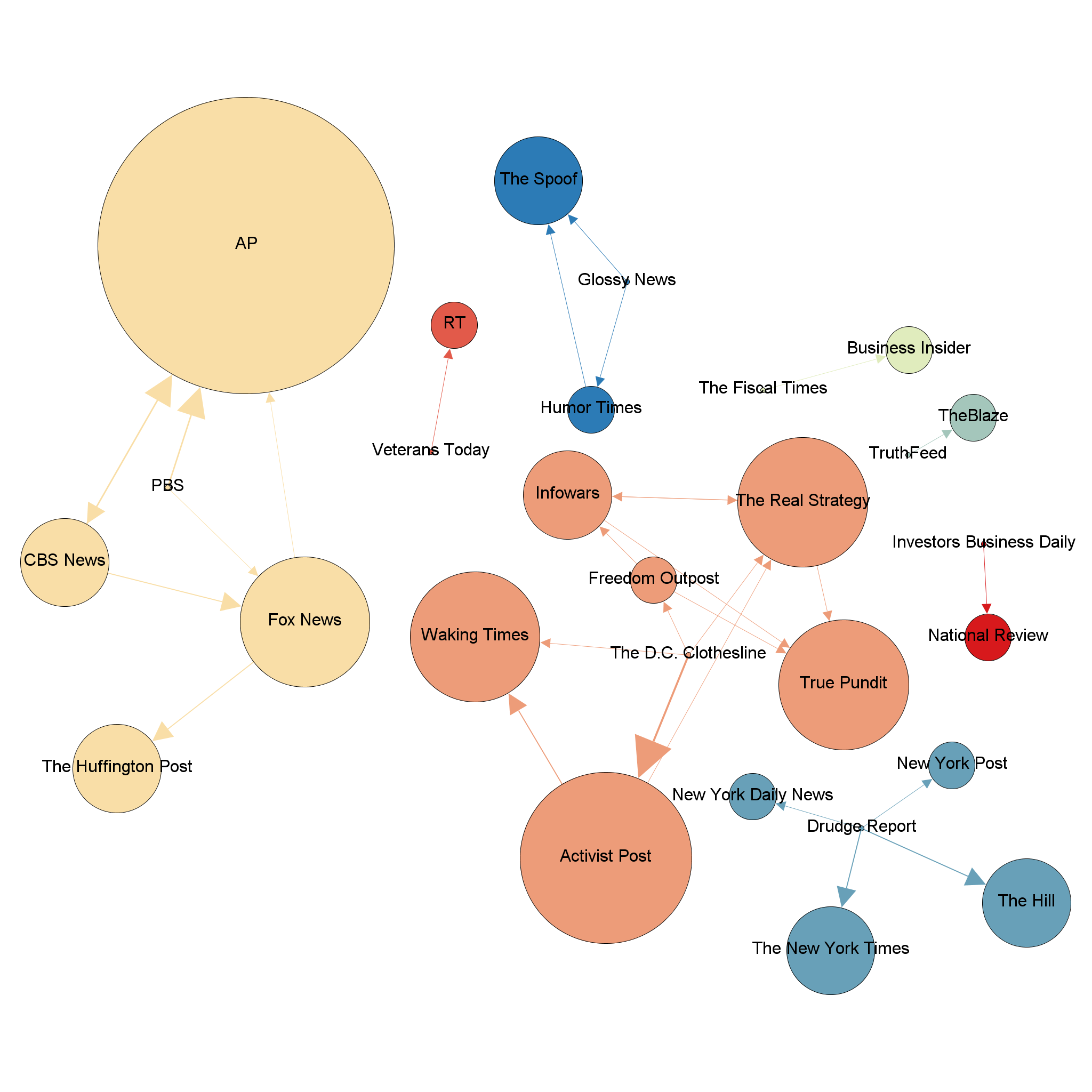}
\end{tabular}
 \vspace*{-0.2in}  \caption{Article similarity graphs during two different two-week periods. The weighted in-degree is the number of articles copied from a source. The weight is indicated by the size of the arrow. The in-degree of a source is shown by the size of the node. The color of a node indicates the community it belongs to based on modularity.} 
\label{attrib_nets}
\end{figure*}

\subsection{Engagement Characterization}
While the NELA2017 data set does not contain labels, such as which articles are fake and which are not, we can make labeled subgroups of the data set using external labeling or unsupervised clustering over different features described in the previous use case.
For space reasons, we provide an example of external labeling only.
There are many ways to label news articles and sources in the NELA2017 data set such as based on ownership, self-proclaimed political leaning, reliability (using a lexicon like \url{opensources.co}), or the age of the news source.  

To explore this method, we group sources by their self-proclaimed political leaning as conservative or liberal and exclude satire news sources and any news source that does not clearly claim a political ideology. These subgroups contain 16 liberal sources and 17 conservative sources. While there are certainly other politically biased news sources in the data set, we are strictly looking at self-proclaimed leaning. We can break down these groups even further by using previously known reporting behavior. Specifically, we ask ``has the source published a completely false article in the past?'' To do this, we manually use 3 online fact-checkers: (\url{snopes.com}, \url{politifact.com} or \url{factcheck.org}). In this division, we do not include sources that have published partially false articles, only completely false. This labeling can be thought of as source-level reliability rather than article-level correctness.

With these newly labeled subgroups of the NELA2017 data set, we explore Facebook shares over time. In Figure~\ref{fb}a, we see that, on average, politically-left leaning news sources had higher shares over the 7 month time period and these shares increased over time. When looking at the max number of shares, rather than the median, we see politically-right leaning news sources were often shared slightly more. In Figure~\ref{fb}b, when splitting by previously publishing a false article, false politically-left sources were shared more than true politically-left news sources in the first 3 months of the time slice, but decrease significantly in the last 4 months of the time slice. In contrast, false right-leaning sources are shared more than true right-leaning source over the full 7 month time slice. While this simple analysis does not conclude that false news articles were more highly shared than true news articles during this time, it does illustrate differences in engagement with political news sources that have published false articles in the past. 


\subsection{Attribution and Content Copying}
A lesser studied area that can benefit from the NELA2017 data set is news attribution, which has been studied in journalism, but not in the context of today's news ecosystem. In context of today's news environment, Jane Lytvynenko of Buzzfeed News points out that the conspiracy news site Infowars copied 1000's are articles from other sources without attribution over the past 3 years~\cite{lytvynenko}. Most notably, Infowars copied from Russia Today (RT), Sputnik, CNN, BBC, The New York Times, Breitbart, CNS News, and The Washington Post. This article sheds light on the potential content-mixing methods of fake and conspiracy news sources that publish original material
with a specific message and also report ``real'' content from other sources to increase their perceived credibility. 

To provide an  example of this, we extract highly similar articles from several two-week intervals. We do this using the cosine similarity between TFIDF (Term-Frequency Inverse Document-Frequency) article vectors, a standard technique in information retrieval. For every article pair from a different source, if the cosine similarity is above 0.90 (meaning the articles are almost verbatim), we extract the article pair and compare time stamps to see which source published the article first. Over each two week interval, we use the time stamp comparison to create a weighted directed graph, in which in-degree is how many articles are copied from the node and out-degree is how many articles a node copies. In Figure~\ref{attrib_nets}, we show networks from two time frames: May 1st-14th and July 1st-14th. In each figure, the weighted in-degree is represented by the size of the arrow. Each node's in-degree is shown by the size of the node and each node is colored based on the community it belongs to (using modularity). Note, since this is a pair-wise analysis, there may be redundant links if the same story is copied by many sources. For example, if several sources copy a story from AP, the network will not only point to AP, but also to the sources that published that story earlier than another source. While there are many potential types of content copying, this analysis is only exploring near exact content copying. Specifically, sources that may mix false and true content would not be captured by the high cosine similarity.

In each graph, there are multiple connected components and clear communities of who copies from who. In particular, we see well-known mainstream sources copy from each other (primarily from AP, a news wire service) and known conspiracy sources copy from each other. In some cases, these two communities are completely disconnected and other times there is a path between them. For example, in Figure~\ref{attrib_nets}a, there exists a path between USA Politics Now and Fox News (through Liberty Writers and The Gateway Pundit). In other time slices (not shown), we see a direct path between Infowars and Fox News (Fox News copying from Infowars and vice versa). In addition to these two larger communities, we see many separate smaller communities of sources, including satire, left-wing, and right-wing communities. We see very similar community structure and attribution patterns throughout the data set. Overall, the community structure we observe in content similarity networks is very similar to that of the news ecosystem on Twitter~\cite{starbird2017examining}, where alternative news sources form tight-knit communities with few connections to mainstream news. 

We further categorize the types of content copying we see into 
three primary categories:

{\bf Proper Attribution, Different Title.} 
Many sources publish full, word-for-word articles from The Associated Press (AP), but provide clear citations such as ``2017 The Associated Press. All Rights Reserved." or ``The Associated Press contributed to this report." Specifically, we see this citation behavior in sources like CBS News, PBS News, Fox News, Breitbart, The Talking Points Memo, and The Huffington Post. More interestingly, while the content is almost exactly the same, the titles can be very different. For example, the title for an AP article was ``Scholars White Houses name gaffe not helping US-China ties," where as the Fox News title for the same article was ``Chinese scholars rip White House staff after name mix up." 
Related, we see that True Pundit directly copies many full articles from The Daily Caller (60 copied articles between April 14th and May 14th). At the end of each article The Daily Caller writes: ``Content created by The Daily Caller News Foundation is available without charge to any eligible news publisher that can provide a large audience.'' Thus, True Pundit's  copying can be considered legitimate attribution. Infowars similarly takes articles from the Daily Caller. 

{\bf Same Author, Different Source.}
Surprisingly, we find the majority of highly similar articles are written by the same author on different sources. There are many examples of this behavior. We see The D.C. Clothesline and Freedom Outpost commonly publish articles written by Tim Brown. The D.C. Clothesline also has articles written by Jay Syrmopoulos, who writes for Activist Post and The Free Thought Project. The Daily Caller, Infowars, and The Real Strategy all have word for word identical articles written by Luke Rosiak. The Waking Times and Activist Post have articles written by Alex Pietrowski. Salon and Media Matters for America have multiple articles written by Cydney Hargis. In satire news, Rodger Freed writes the same articles for The Spoof, Humor Times, and Glossy News, usually publishing on The Spoof first. 
In another example, a series of stories about a ``George Soros backed Trump resistance fund'' are published word for word on both Infowars and Fox News, all written by Joe Schoffstal. Each article does not have clear attribution to one or the other source, despite being exact copies and each article was written on Infowars days prior to its publication on Fox News. This example is particularly surprising as Fox News captures a wide, mainstream audience and Infowars is a well known conspiracy source, creating a clear path between a well-established news source and conspiracy/false news. 
Note, while many of these articles are clearly written by the same author, as the authors state they contribute to both sources, there are others that may just be copied with the authors name included. For example, The D.C. Clothesline seems to have many authors that contribute elsewhere, but there is no indication in the authors' biographical information (on the other sources they contribute to) that they contribute to The D.C. Clothesline. Hence, while the author contributes to multiple sources, it is unclear that they contribute to The D.C. Clothesline.
  
{\bf No Attribution.}
We also see several sources, particularly those who have been caught spreading false news in the past, copying news articles with no citation. In particular, we found that both Veterans Today and Infowars copied multiple articles directly from Russia Today (RT),  with no citation similar to behavior
that has been pointed out by Jane Lytvynenko~\cite{lytvynenko}. 

\subsection{Issue framing and narrative slant}
In addition to ``big picture'' analysis, NELA2017 can also be used to study specific events. To illustrate this, we explore differing narratives reported around a specific event. While many sources may cover the same topic, they may not report all sides of a story or may have an imbalanced quantity of coverage~\cite{lin2011more}. This type of coverage bias has been explored in terms of political party slant in US congress stories~\cite{lin2011more}, and similar notions of bias, including framing and agenda setting bias, have been in explored in various media studies~\cite{entman2007framing}~\cite{pan1993framing}. There is more recent work on ideological bias in news stories caused by journalists Twitter networks~\cite{wihbey2017exploring}. However, there is little to no recent work on the specific dynamics of differing news narratives. Further, since the NELA2017 data set covers many different political events, it is ideal for tracking publishing and reporting behavior over a wide range of time, something that has also not been explored in the literature.

To provide an example of event extraction from the NELA2017 data set, we perform a simple extraction technique on two different events: \begin{enumerate*} \item the U.S. national anthem protests~\footnote{\url{en.wikipedia.org/wiki/U.S._national_anthem_protests}}, \item the dismissal of James Comey~\footnote{\url{en.wikipedia.org/wiki/Dismissal_of_James_Comey}}\end{enumerate*}. The U.S. national anthem protests were protests in which athletes, specifically NFL players, kneeled during the singing of the U.S. national anthem to protest police brutality and racial inequality in the U.S. These protests begin in 2016, but became widespread in late 2017 as U.S. President Donald Trump called for NFL team owners to fire any player who kneeled. This event caused a debate of whether NFL players were being disrespectful to the U.S. flag and military. Hence, two sides of the story emerged: race inequality and disrespecting the military. A similar two-sided story is the dismissal of James Comey. James Comey was the 7th director of the Federal Bureau of Investigation (FBI), who was dismissed by U.S. President Donald Trump in May 2017. This dismissal came at a controversial time, as President Trump's administration was under FBI investigation for alleged Russian interference in the 2016 election. At the same time, James Comey 
had been widely criticized for the way he handled the earlier Hilary Clinton email controversy~\footnote{\url{en.wikipedia.org/wiki/Hillary_Clinton_email_controversy}}. The Trump administration publicly stated Comey's dismissal was due to the recommendation by then 
Attorney General Jeff Sessions and Comey's handling of the earlier email investigation. The media created a divide between the two sides: did President Trump dismiss Comey due to the Russia investigation or due to the Clinton email investigation. Therefore, in both of these events there are clear sides that news sources may or may not give fair coverage. 

To do this analysis, we first select the dates of each event and extract all articles from several days before and after the event. With these articles extracted, we filter by a set of event keywords and manually ensure all articles extracted are reporting the appropriate event. We then modify a simple slant score technique used in~\cite{lin2011more} to quantify the narrative slant. In ~\cite{lin2011more}, the slant score is measured by the log-odds-ratio of the number of times source $i$ refers to party $k$ (specifically refers to a member of said party), where the baseline probability is 50\% (meaning an article has a 50-50 chance to refer to each party). We perform a similar analysis, but instead of counting party references, we count narrative keyword references. These narrative keywords are manually generated. While there are more sophisticated methods to measure bias, this method provides a base understanding of coverage bias within these stories.

\begin{figure*}
\begin{center}
 \hspace*{-0.1in}\begin{tabular}{cccc} \\
  \multicolumn{2}{c}{(a) NFL Protests (Sept 20th 2017 to Sept Sept 30th 2017)} &     
  \multicolumn{2}{c}{(b) Comey Firing (May 10th 2017 to May 15th 2017)} \\
    \includegraphics[width=115pt,keepaspectratio=true]{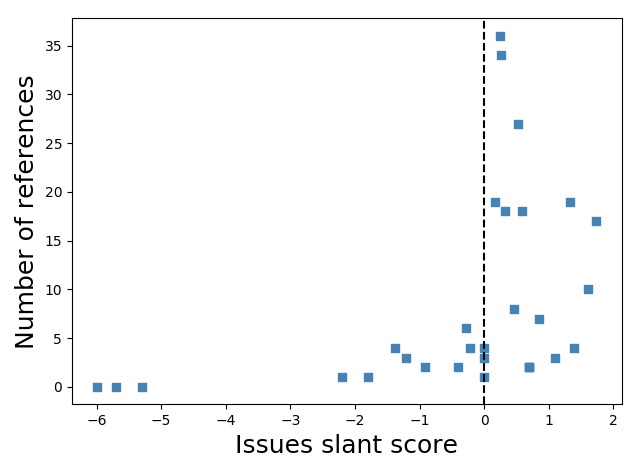}
    & \includegraphics[width=115pt,keepaspectratio=true]{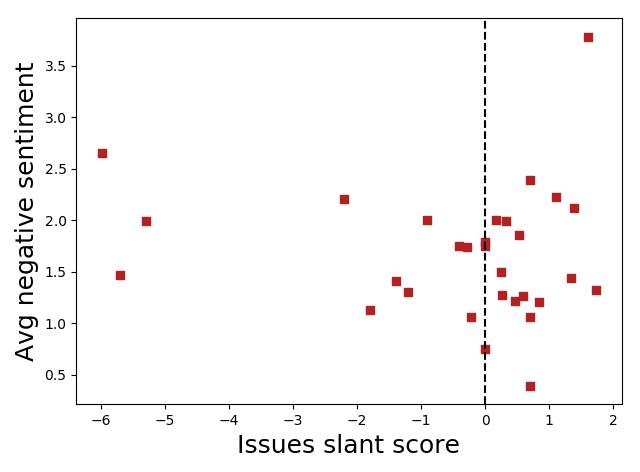}    
    & 
    \includegraphics[width=115pt,keepaspectratio=true]{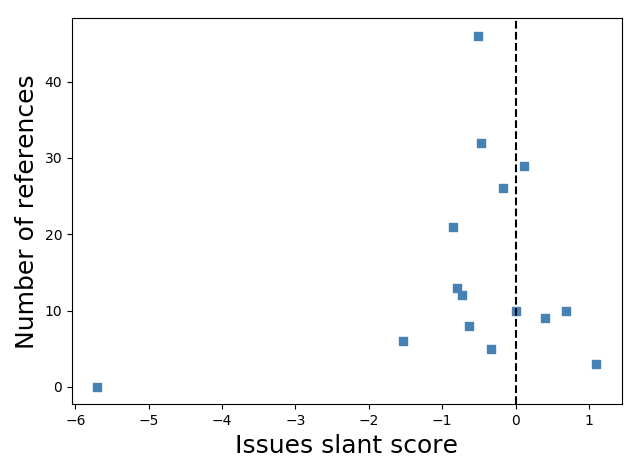}
    & \includegraphics[width=120pt,keepaspectratio=true]{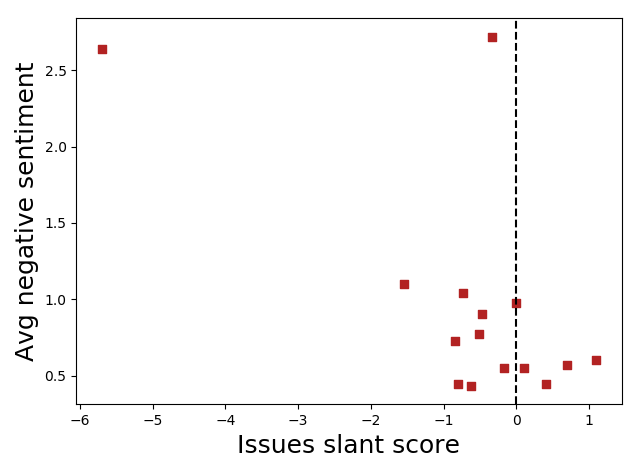}    \\
    
     \includegraphics[width=115pt,keepaspectratio=true]{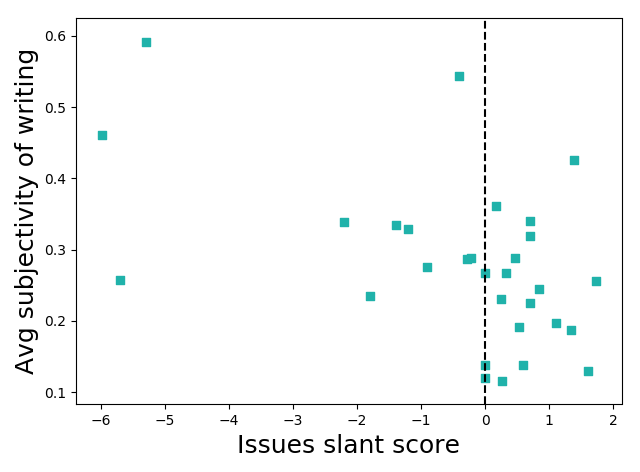}    
    & \includegraphics[width=115pt,keepaspectratio=true]{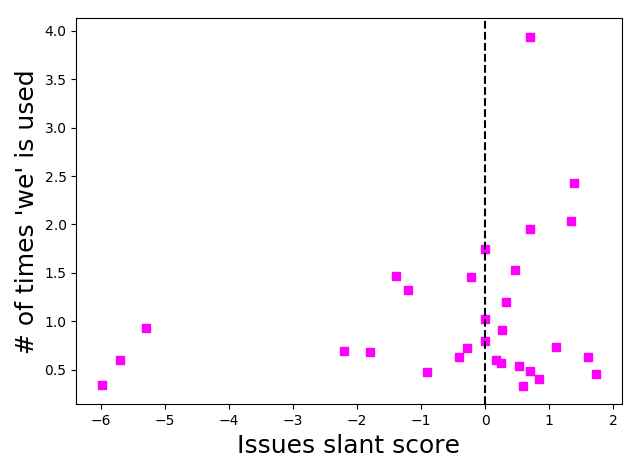}    
    & 
     \includegraphics[width=115pt,keepaspectratio=true]{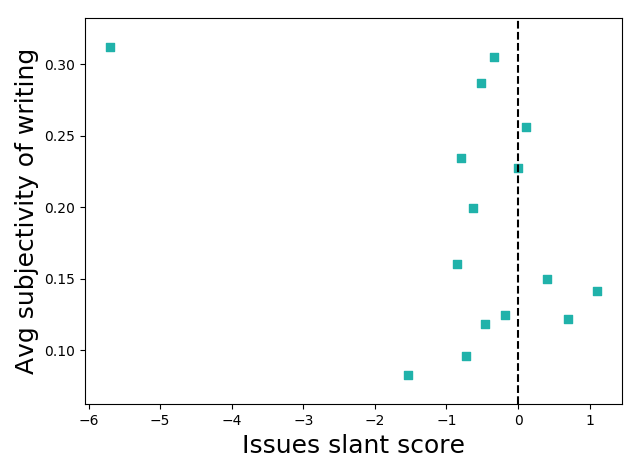}    
    & \includegraphics[width=115pt,keepaspectratio=true]{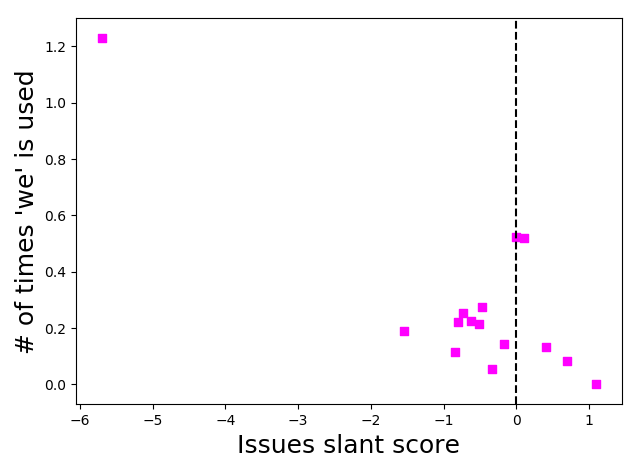}    \\
\end{tabular}
\vspace*{-0.2in}  \caption{Issue slant score computed using the log-odds-ratio of narrative keywords. Each dot in the scatter plot represents a source, the x-axis the slant score, and the y-axis is the overall number of references or a feature from the NELA2017 feature set. A score of 0 indicated by the vertical line is perfectly balanced coverage. In (a) sources with higher scores report more about what the players are protesting (police brutality) than the disrespect for the flag and military (lower score is vice-versa). In (b) sources with higher scores report more about the Russia collusion than the Clinton email scandal (lower score is vice-versa). \label{slants}}
\end{center}
\end{figure*}

{\bf U.S. national anthem protests.}
For the U.S. national anthem protests, we use the following keywords for side 1: \textit{ Kaepernick, racism, race, racist, police, brutality, African American, and prejudice}, and the following for side 2: \textit{respect, stand, disrespect, flag, troops, military}, and \textit{anti-American}.

In Figure~\ref{slants}a, we show a scatter plot in which each point represents a source and the x-axis shows the computed slant score. If a source reported both sides equally, it receives a slant score of 0 (indicated by the vertical dotted line). In this case, the higher the score the more coverage of side 1 (police brutality) and the lower the score the more coverage of side 2 (disrespect of flag). On the y-axis we show the either the number of keyword references overall or a feature selected from the NELA2017 feature set. 

We can see right away that there are sources across the spectrum of coverage slant; however, more so on side 1 (police brutality). Despite more sources covering side 1, we see more extreme slant (further from balanced) for side 2 (disrespect of flag), meaning they mention keywords corresponding to side 2 much more than side 1. When inspecting the sources with this extreme slant, we see several cases where there was no mention of side 1. Whereas even the most extreme slant towards side 2 mentions the debate of respecting the flag. Of those sources that only report the disrespecting of the flag narrative, we see they are more subjective in writing and slightly more negative than those sources who are near balanced. On the other side, those who report more of the police brutality message use the 1st person plural words more (like we, us, or our). 

{\bf Dismissal of James Comey}
For the dismissal of James Comey, we use the following keywords for side 1: \textit{Russia, Trump-Russia, collusion, election}, and \textit{meddling}, and the following for side 2: \textit{Hilary, Clinton, Democrats, dems, email}, and \textit{server}. In Figure~\ref{slants}b, we show the same for scatter plots as in Figure~\ref{slants}a discussed above. In this case, the higher the score the more coverage of side 1 (Russia) and the lower the score the more coverage of side 2 (Clinton emails).

In this story, we can see the vast majority of sources give balanced coverage, receiving slant scores close to 0. In fact, there is only 1 source that reported the event extremely one sided. When inspecting this source, they did not mention anything about the Russia investigation, only the Clinton email scandal. This one extreme source was much more negative, more subjective, and used 1st person plurals more than the other sources. 

\section{Conclusions}
In this paper, we presented the NELA2017 data set, which contains articles from 92 news sources over 7 months, as well as, 130 content-based 
features that have been used throughout the news literature. Together with the data set,
we include the source code for computing the features (\url{goo.gl/JsxGt}). 
We also illustrated potential 
research directions with a number of use cases, showing the data set's use in studying
individual sources, compare sources to each other, or study sources over a specific event. We are continuing to expand and collect data for future releases. As we update, we will release the data set by versions, thus, NELA2017 will be an unchanged version corresponding to the meta data in this paper. All data can be requested at \url{nelatoolkit.science}.

\bibliographystyle{aaai}

\begin{thebibliography}{}

\bibitem[\protect\citeauthoryear{An, Aldarbesti, and
  Kwak}{2017}]{an2017convergence}
An, J.; Aldarbesti, H.; and Kwak, H.
\newblock 2017.
\newblock Convergence of media attention across 129 countries.
\newblock In {\em Intl. Conf. on Social Informatics},  159--168.
\newblock Springer.

\bibitem[\protect\citeauthoryear{Bessi \bgroup et al\mbox.\egroup
  }{2015}]{Bessi:2015hg}
Bessi, A.; Coletto, M.; Davidescu, G.~A.; Scala, A.; Caldarelli, G.; and
  Quattrociocchi, W.
\newblock 2015.
\newblock {Science vs Conspiracy: Collective Narratives in the Age of
  Misinformation}.
\newblock {\em PLoS ONE} 10(2):e0118093--17.

\bibitem[\protect\citeauthoryear{Buntain and
  Golbeck}{2017}]{buntain2017automatically}
Buntain, C., and Golbeck, J.
\newblock 2017.
\newblock Automatically identifying fake news in popular twitter threads.
\newblock In {\em 2017 IEEE Intl. Conference on Smart Cloud},  208--215.
\newblock IEEE.

\bibitem[\protect\citeauthoryear{Chakraborty \bgroup et al\mbox.\egroup
  }{2016}]{chakraborty2016stop}
Chakraborty, A.; Paranjape, B.; Kakarla, S.; and Ganguly, N.
\newblock 2016.
\newblock Stop clickbait: Detecting and preventing clickbaits in online news
  media.
\newblock In {\em ASONAM},  9--16.
\newblock IEEE.

\bibitem[\protect\citeauthoryear{Entman}{2007}]{entman2007framing}
Entman, R.~M.
\newblock 2007.
\newblock Framing bias: Media in the distribution of power.
\newblock {\em Journal of communication} 57(1):163--173.

\bibitem[\protect\citeauthoryear{Graham, Haidt, and
  Nosek}{2009}]{graham2009liberals}
Graham, J.; Haidt, J.; and Nosek, B.~A.
\newblock 2009.
\newblock Liberals and conservatives rely on different sets of moral
  foundations.
\newblock {\em Journal of personality and social psychology} 96(5):1029.

\bibitem[\protect\citeauthoryear{Horne, Adali, and
  Sikdar}{2017}]{horne2017identifying}
Horne, B.~D.; Adali, S.; and Sikdar, S.
\newblock 2017.
\newblock Identifying the social signals that drive online discussions: A case
  study of reddit communities.
\newblock In {\em Computer Communication and Networks (ICCCN), 2017 26th
  International Conference on},  1--9.
\newblock IEEE.

\bibitem[\protect\citeauthoryear{Horne and Adal{\i}}{2017}]{horne2017just}
Horne, B.~D., and Adal{\i}, S.
\newblock 2017.
\newblock This just in: Fake news packs a lot in title, uses simpler,
  repetitive content in text body, more similar to satire than real news.
\newblock In {\em ICWSM NECO Workshop}.

\bibitem[\protect\citeauthoryear{Horne \bgroup et al\mbox.\egroup
  }{2018}]{horne2018accessing}
Horne, B.~D.; Dron, W.; Khedr, S.; and Adali, S.
\newblock 2018.
\newblock Assessing the news landscape: A multi-module toolkit for evaluating
  the credibility of news.
\newblock In {\em WWW Companion}.

\bibitem[\protect\citeauthoryear{Hutto and Gilbert}{2014}]{hutto2014vader}
Hutto, C.~J., and Gilbert, E.
\newblock 2014.
\newblock Vader: A parsimonious rule-based model for sentiment analysis of
  social media text.
\newblock In {\em ICWSM}.

\bibitem[\protect\citeauthoryear{Kwak and An}{2016}]{kwak2016revealing}
Kwak, H., and An, J.
\newblock 2016.
\newblock Revealing the hidden patterns of news photos: Analysis of millions of
  news photos through gdelt and deep learning-based vision apis.
\newblock In {\em ICWSM}.

\bibitem[\protect\citeauthoryear{Lin, Bagrow, and Lazer}{2011}]{lin2011more}
Lin, Y.-R.; Bagrow, J.~P.; and Lazer, D.
\newblock 2011.
\newblock More voices than ever? quantifying media bias in networks.
\newblock {\em ICWSM} 1(arXiv: 1111.1227):1.

\bibitem[\protect\citeauthoryear{Lin \bgroup et al\mbox.\egroup
  }{2017}]{lin2017acquiring}
Lin, Y.; Hoover, J.; Dehghani, M.; Mooijman, M.; and Ji, H.
\newblock 2017.
\newblock Acquiring background knowledge to improve moral value prediction.
\newblock {\em arXiv preprint arXiv:1709.05467}.

\bibitem[\protect\citeauthoryear{Loper and Bird}{2002}]{loper2002nltk}
Loper, E., and Bird, S.
\newblock 2002.
\newblock Nltk: The natural language toolkit.
\newblock In {\em ACL-02 Workshop on Effective tools and methodologies for
  teaching natural language processing and computational linguistics},  63--70.

\bibitem[\protect\citeauthoryear{Lytvynenko}{2017}]{lytvynenko}
Lytvynenko, J.
\newblock 2017.
\newblock Infowars has republished more than 1,000 articles from rt without
  permission.
\newblock Available at
  ``\url{www.buzzfeed.com/janelytvynenko/infowars-is-running-rt-content}".

\bibitem[\protect\citeauthoryear{Mele \bgroup et al\mbox.\egroup
  }{2017}]{mele2017combating}
Mele, N.; Lazer, D.; Baum, M.; Grinberg, N.; Friedland, L.; Joseph, K.; Hobbs,
  W.; and Mattsson, C.
\newblock 2017.
\newblock Combating fake news: An agenda for research and action.

\bibitem[\protect\citeauthoryear{Mitchell \bgroup et al\mbox.\egroup
  }{2013}]{mitchell2013geography}
Mitchell, L.; Frank, M.~R.; Harris, K.~D.; Dodds, P.~S.; and Danforth, C.~M.
\newblock 2013.
\newblock The geography of happiness: Connecting twitter sentiment and
  expression, demographics, and objective characteristics of place.
\newblock {\em PloS one} 8(5):e64417.

\bibitem[\protect\citeauthoryear{Mitra and Gilbert}{2015}]{mitra2015credbank}
Mitra, T., and Gilbert, E.
\newblock 2015.
\newblock Credbank: A large-scale social media corpus with associated
  credibility annotations.
\newblock In {\em ICWSM},  258--267.

\bibitem[\protect\citeauthoryear{Mukherjee and
  Weikum}{2015}]{mukherjee2015leveraging}
Mukherjee, S., and Weikum, G.
\newblock 2015.
\newblock Leveraging joint interactions for credibility analysis in news
  communities.
\newblock In {\em CIKM},  353--362.
\newblock ACM.

\bibitem[\protect\citeauthoryear{Pan and Kosicki}{1993}]{pan1993framing}
Pan, Z., and Kosicki, G.~M.
\newblock 1993.
\newblock Framing analysis: An approach to news discourse.
\newblock {\em Political communication} 10(1):55--75.

\bibitem[\protect\citeauthoryear{Pang and Lee}{2004}]{pang2004sentimental}
Pang, B., and Lee, L.
\newblock 2004.
\newblock A sentimental education: Sentiment analysis using subjectivity
  summarization based on minimum cuts.
\newblock In {\em ACL},  271.

\bibitem[\protect\citeauthoryear{Piotrkowicz \bgroup et al\mbox.\egroup
  }{2017}]{piotrkowicz2017headlines}
Piotrkowicz, A.; Dimitrova, V.; Otterbacher, J.; and Markert, K.
\newblock 2017.
\newblock Headlines matter: Using headlines to predict the popularity of news
  articles on twitter and facebook.
\newblock In {\em ICWSM},  656--659.

\bibitem[\protect\citeauthoryear{Popat \bgroup et al\mbox.\egroup
  }{2016}]{popat2016credibility}
Popat, K.; Mukherjee, S.; Str{\"o}tgen, J.; and Weikum, G.
\newblock 2016.
\newblock Credibility assessment of textual claims on the web.
\newblock In {\em CIKM},  2173--2178.
\newblock ACM.

\bibitem[\protect\citeauthoryear{Potthast \bgroup et al\mbox.\egroup
  }{2017}]{potthast2017stylometric}
Potthast, M.; Kiesel, J.; Reinartz, K.; Bevendorff, J.; and Stein, B.
\newblock 2017.
\newblock A stylometric inquiry into hyperpartisan and fake news.
\newblock {\em arXiv preprint arXiv:1702.05638}.

\bibitem[\protect\citeauthoryear{Qian and Zhai}{2014}]{qian2014unsupervised}
Qian, M., and Zhai, C.
\newblock 2014.
\newblock Unsupervised feature selection for multi-view clustering on
  text-image web news data.
\newblock In {\em CIKM},  1963--1966.
\newblock ACM.

\bibitem[\protect\citeauthoryear{Recasens, Danescu-Niculescu-Mizil, and
  Jurafsky}{2013}]{recasens2013linguistic}
Recasens, M.; Danescu-Niculescu-Mizil, C.; and Jurafsky, D.
\newblock 2013.
\newblock Linguistic models for analyzing and detecting biased language.
\newblock In {\em ACL (1)},  1650--1659.

\bibitem[\protect\citeauthoryear{Reis \bgroup et al\mbox.\egroup
  }{2015}]{reis2015breaking}
Reis, J.; Benevenuto, F.; de~Melo, P.~V.; Prates, R.; Kwak, H.; and An, J.
\newblock 2015.
\newblock Breaking the news: First impressions matter on online news.
\newblock In {\em ICWSM}.

\bibitem[\protect\citeauthoryear{Saez-Trumper, Castillo, and
  Lalmas}{2013}]{saez2013social}
Saez-Trumper, D.; Castillo, C.; and Lalmas, M.
\newblock 2013.
\newblock Social media news communities: gatekeeping, coverage, and statement
  bias.
\newblock In {\em CIKM},  1679--1684.
\newblock ACM.

\bibitem[\protect\citeauthoryear{Shao \bgroup et al\mbox.\egroup
  }{2016}]{Shao:2016:HPT:2872518.2890098}
Shao, C.; Ciampaglia, G.~L.; Flammini, A.; and Menczer, F.
\newblock 2016.
\newblock Hoaxy: A platform for tracking online misinformation.
\newblock In {\em WWW 2016},  745--750.

\bibitem[\protect\citeauthoryear{Singhania, Fernandez, and
  Rao}{2017}]{singhania20173han}
Singhania, S.; Fernandez, N.; and Rao, S.
\newblock 2017.
\newblock 3han: A deep neural network for fake news detection.
\newblock In {\em International Conference on Neural Information Processing},
  572--581.
\newblock Springer.

\bibitem[\protect\citeauthoryear{Starbird}{2017}]{starbird2017examining}
Starbird, K.
\newblock 2017.
\newblock Examining the alternative media ecosystem through the production of
  alternative narratives of mass shooting events on twitter.
\newblock In {\em ICWSM},  230--239.

\bibitem[\protect\citeauthoryear{Wang}{2017}]{wang2017liar}
Wang, W.~Y.
\newblock 2017.
\newblock " liar, liar pants on fire": A new benchmark dataset for fake news
  detection.
\newblock {\em arXiv preprint arXiv:1705.00648}.

\bibitem[\protect\citeauthoryear{Wihbey \bgroup et al\mbox.\egroup
  }{2017}]{wihbey2017exploring}
Wihbey, J.; Coleman, T.~D.; Joseph, K.; and Lazer, D.
\newblock 2017.
\newblock Exploring the ideological nature of journalists' social networks on
  twitter and associations with news story content.
\newblock {\em arXiv preprint arXiv:1708.06727}.

\bibitem[\protect\citeauthoryear{Zubiaga \bgroup et al\mbox.\egroup
  }{2016}]{zubiaga2016analysing}
Zubiaga, A.; Liakata, M.; Procter, R.; Hoi, G. W.~S.; and Tolmie, P.
\newblock 2016.
\newblock Analysing how people orient to and spread rumours in social media by
  looking at conversational threads.
\newblock {\em PloS one} 11(3):e0150989.

\end{thebibliography}

\end{document}